\documentclass{article}
\usepackage{epsf}
\newcommand{\mean}[1]{\mbox{$\left<#1\right>$}}

\title{
Monte Carlo simulations of magnetovolume instabilities in
anti-Invar systems}
\author{M.E.\ Gruner and P.\ Entel\\[1ex] \normalsize
  Theoretische Tieftemperaturphysik \\ \normalsize
  Gerhard-Mercator-Universit\"at -- GH Duisburg, D-47048 Duisburg,
  Germany\\ \normalsize
  Tel.: +49-203-379-3564,~ Fax.: +49-203-379-2965,~
  eMail: me@thp.Uni-Duisburg.DE\\[2ex] \normalsize
  Presented at ICAM '97 / E-MRS spring meeting, Strasbourg,
  France, 1997 } %\\ \normalsize  to appear in Comp. Mat. Sci. (1998)}

\begin{document}
\renewcommand{\baselinestretch}{1.0}
\maketitle
\subsection*{\normalsize\bf\centering Abstract}
We perform constant pressure Monte Carlo simulations of a
spin-analogous model which describes coupled spatial and magnetic
degrees of freedom on an fcc lattice.
Our calculations qualitatively reproduce magnetovolume effects
observed in some rare earth manganese compounds, especially in
the anti-Invar material ${\rm YMn}_2$.
These are a sudden collapse of the magnetic moment
which is connected with a huge volume change, and a
largely enhanced thermal expansion coefficient.

\section*{\normalsize\bf INTRODUCTION}
The observation that some transition metal alloys and complexes have
an anomalously low or high thermal expansion coefficient is widely
referred to as the Invar or anti-Invar effect, respectively.
Since the discovery of Fe-Ni Invar one century ago,
uncounted investigations in this topic have
taken place (for recent reviews see \cite{Wassermann}).
A major advance has been achieved by ab initio
band structure calculations (e.\ g.\ \cite{Entel,Janne}).
KKR-CPA calculations of ${\rm Fe}_{65}{\rm Ni}_{35}$
\cite{Janne} reveal two nearly degenerate minima on the
binding surface,
i.e.\ the energy as a function of lattice constant and mean magnetic
moment. One at a large lattice constant and a large magnetic moment
(high moment or HM state) and another at a smaller lattice constant
and a smaller or vanishing magnetic moment (low moment or LM state).
In this case,
thermal excitation of the LM state can lead to a compensation of the volume
expansion of the material.
This scenario reminds of the phenomenological 2-$\gamma$-states model
\cite{Weiss} introduced by Weiss in order to explain Fe-Ni Invar.
However,
most ab initio calculations are performed for zero
temperature only. So
the comparison with experimental
data is hampered by the fact that the
extrapolation of these results to finite temperatures is connected
with severe methodological problems.
So far mainly continuous Ginzburg Landau like spin models in
combination with a Gaussian fluctuation theory have been used to
achieve this goal \cite{GL}. Another approach is to use a
Weiss-type model with its parameters adapted to ab initio results.
This has been done for the classical Invar system Fe-Ni, and
although we neglected the itinerant character of magnetism by using
localized spins, we were
able to reproduce qualitatively the major magnetovolume effects of
this alloy \cite{FeNiPaper}.

However, it has still to be proven, whether this approach can be
applied to other materials, especially to anti-Invar
systems. One of the most prominent anti-Invar materials is
the Laves phase compound ${\rm YMn}_2$. It occurs in the C15 structure
with a frustrated antiferromagnetic (AF) spin order in the Mn
sublattice.
Around $T_{\rm N}\approx100\;{\rm K}$ ${\rm YMn}_2$ undergoes a first order
phase transition into a paramagnetic state,
accompanied by a huge volume contraction of $5\:\%$ and a large
thermal hysteresis of $30\;{\rm K}$. Above $T_{\rm N}$ a large
thermal expansion coefficient of $50\times10^{-6}\;{\rm K}^{-1}$
is encountered \cite{YMn2}.
LMTO calculations reveal a nonmagnetic solution at a
lower volume with a slightly higher energy than the AF ground state
\cite{Asano}.

Within the scope of this work, we will show that for a suitable choice
of parameters the large abrupt volume change at $T_{\rm N}$ as well
as the enhanced thermal expansion of ${\rm YMn}_2$ can be
qualitatively reproduced by a Weiss type model assuming a HM-LM
instability of the Mn atom.

\section*{\normalsize\bf THE MODEL}
Our model Hamiltonian consists of a magnetic part $H_{\rm m}$
describing the the magnetic properties of the
system and a vibrational part $H_{\rm v}$ responsible for
the magnetovolume coupling:
\begin{equation}
  H = \underbrace{D \sum_i S_i^2
    + J \!\!\sum_{<i,k>}\!\! S_i S_k}_{H_{\rm m}}
  + \!\!\underbrace{\sum_{<i,k>}\!\!
    U(r_{ik},S_i,S_k)}_{H_{\rm v}}
  \;\mbox{.}\label{eq:H}
\end{equation}
$H_{\rm m}$
is a spin-1 Ising Hamiltonian, also known as Blume-Capel model. 
We identify the spin states $S_i=0$ with the atomic LM state, whereas
$S_i=\pm 1$ refers to the HM state.
$D$ denotes the crystal field, which imposes an energetic separation
between HM and LM states; $J$ is the magnetic
exchange constant. The brackets indicate a summation over nearest
neighbors. We did not consider a distance dependence of $J$ so far,
since we believe its influence to be small compared to the effect observed.
Furthermore, compressible frustrated Ising antiferromagnets show additional
effects like tetragonal lattice distortions \cite{AFCI}, that are not
covered by experimental data in our case.
The second part $H_{\rm v}$ introduces pair interactions
between the atoms, which depend on their distance $r_{ik}$ and on their
spin states.
If both of the interacting atoms are in the HM state, we employ HM
potentials that are characterized by a larger equilibrium distance than
the LM potentials we use otherwise.
So a system with a considerable concentration of LM atoms is supposed
to have a smaller lattice constant than a pure HM system.
For the sake of simplicity we choose Lennard-Jones interactions:
\begin{equation}
  U(r_{ik},S_i,S_k)=\left\{
    \begin{array}{lcl}
      4\epsilon_{\rm L}\left(\left(\frac{\displaystyle d_{\rm L}}{\displaystyle
        r_{ik}}\right)^{12}\!\!\!-
        \left(\frac{\displaystyle d_{\rm L}}{\displaystyle
        r_{ik}}\right)^{6}\right) ,
      & \mbox{$S_i S_k=0$} \\
      4\epsilon_{\rm H}\left(\left(\frac{\displaystyle d_{\rm H}}{\displaystyle
        r_{ik}}\right)^{12}\!\!\!-
        \left(\frac{\displaystyle d_{\rm H}}{\displaystyle
        r_{ik}}\right)^{6}\right) ,
      &\mbox{$S_i S_k\neq 0$} \\
    \end{array}
  \right.\!\!\mbox{,}
  \label{eq:LJ}
\end{equation}
where $\epsilon_{\rm L,H}$ denote the energy at the equilibrium
neighbor distances $r_{ik}=2^{1/6}d_{\rm L,H}$
for LM and HM potentials, respectively.

\section*{\normalsize\bf THE SIMULATION}
Our Monte Carlo routine consists of two local and one global update steps.
For each lattice site we choose a new spin state $S_i\in\{0,\pm 1\}$ using
the Metropolis criterion.
Afterwards a new trial position is elected out of a cube around the
old position.
Again, the new position is accepted with the probability
$\max(1,\exp(-\beta\Delta H))$, where $\beta$ is the inverse
temperature.
The size of the cube is given by the condition that about half of the
propositions are to be accepted in order to improve convergence.
When all atoms have been updated, the volume of the complete lattice
is adapted by another Metropolis step, except that now the quantity
\mbox{$\Delta{\cal H} = \Delta H - N\:k_{\rm B}\:T\:\ln\left(V'/V\right)$}
has to be considered, which accounts for the difference in translational
entropy caused by the change of the volume from $V$ to $V'$
(for more information on constant pressure Monte Carlo methods see
\cite{AllenTildesley}).
Our calculations were performed on a $8^3\times 4=2048$ site fcc
lattice with periodic boundary conditions at constant pressure $P=0$.
$80\:000$ lattice sweeps
were performed for each temperature. The first $40\:000$ lattice sweeps
were used to allow the system to reach equilibrium, afterwards every
tenth sweep data were gathered for the statistics.

Although ${\rm YMn}_2$ occurs in the C15 structure, we chose instead an fcc
lattice in our simulations. This allows us to omit the Y
species, which we do not expect to play an important role in the
thermodynamics of the HM-LM transition. Furthermore we can easily
relate our results to previous investigations \cite{FeNiPaper}.
Since the fcc Ising antiferromagnet
has also a frustrated ground state spin structure as it is assumed for
${\rm YMn}_2$, simulating an fcc
structure is not a serious restriction.
We used $d_{\rm H}=2.432\;{\rm\AA}$ and $d_{\rm L}=2.397\;{\rm\AA}$
for the interatomic spacing between the Mn 
atoms in AF and paramagnetic ${\rm YMn}_2$.
We chose $\epsilon_{\rm L}=\epsilon_{\rm H}=30.86\;{\rm mRy}$ in order to
achieve a reasonable elastic behavior.
The exchange constant was set to $J=0.459\;{\rm mRy}$ corresponding to
a N\'{e}el temperature $T_{\rm N}=126.5\;{\rm K}$ \cite{Styer}.
The crystal field $D$ was varied in the range $D=0\ldots 3\:J$.
We examined the energy per lattice site \mean{e}, the mean magnetic
moment $\mean{S^2}=\mean{1/N\:\sum_i S_i^2}$, the atomic volume
\mean{V} and the linear thermal expansion coefficient
$\alpha=1/(3V)\:dV/dT$ (brackets denote thermal averaging).
Additionally, for several values of $D$, we approximated ground state
energy and average magnetic moment as a function of the lattice constant
by exponentially cooling down the system to $T=4\;{\rm K}$ at constant
volume.

\section*{\normalsize\bf RESULTS}
\begin{figure} %[p]
\epsfxsize=13cm\epsffile{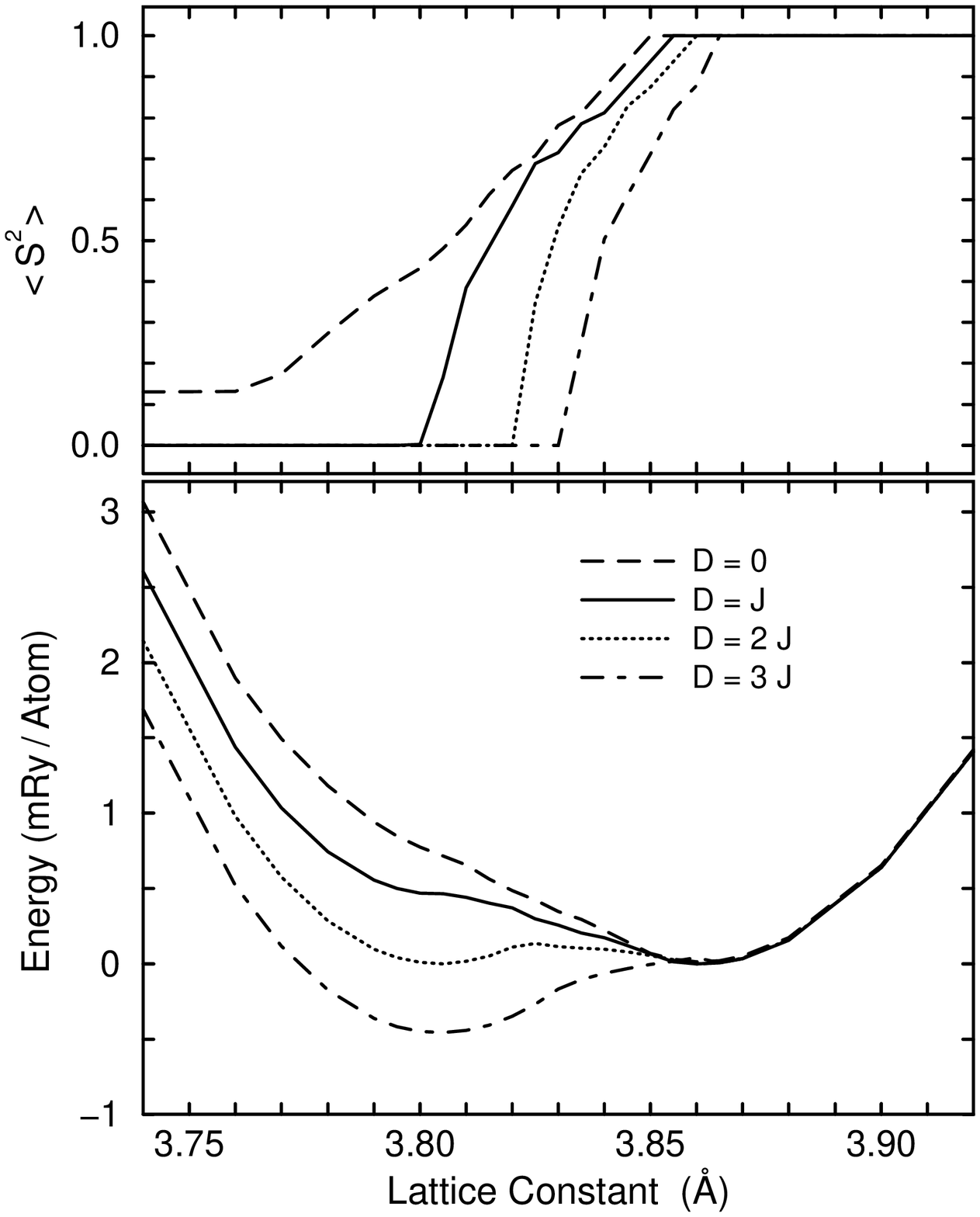}
\caption{Energy per lattice site and average magnetic moment at
  $T=4\;{\rm K}$ as a function of the lattice constant.
  \label{fig:Energy}}
\end{figure}

For $D<2\:J$, the ground state of the system
is an AF HM state as can be seen from simple energetic
considerations concerning Hamiltonian (\ref{eq:H}). 
Correspondingly, for $D>2\:J$, the ground state turns into a
nonmagnetic LM state with a smaller volume.
This is also verified by the plots of low temperature energy and
magnetic moment vs.\ lattice constant shown in Fig.~\ref{fig:Energy}.

\begin{figure} %[p]
\epsfxsize=13cm\epsffile{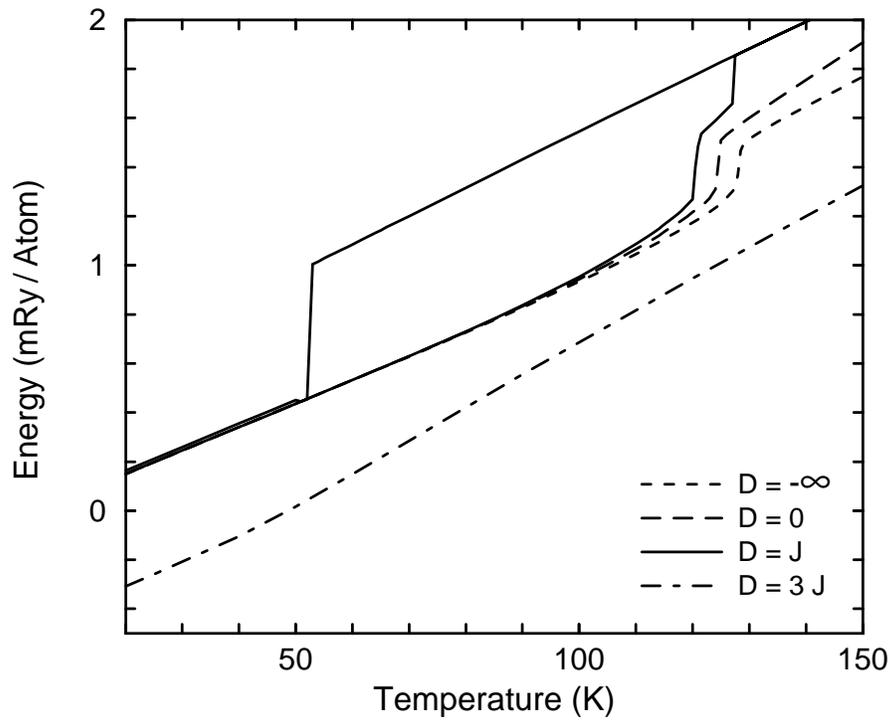}
\caption{Energy per lattice site for various values of $D$ as a
  function of the temperature.
  \label{fig:e}}
\end{figure}

\begin{figure} %[p]
\epsfxsize=13cm\epsffile{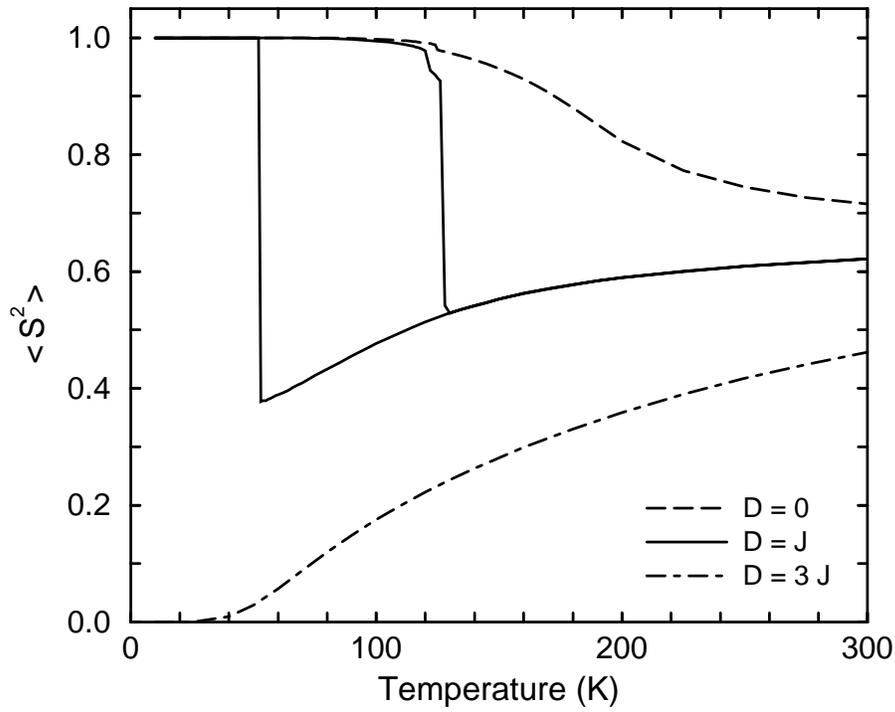}
\caption{Temperature dependence of the average magnetic moment  for
  various values of $D$.
  \label{fig:S2}}
\end{figure}

\begin{figure} %[p]
\epsfxsize=13cm\epsffile{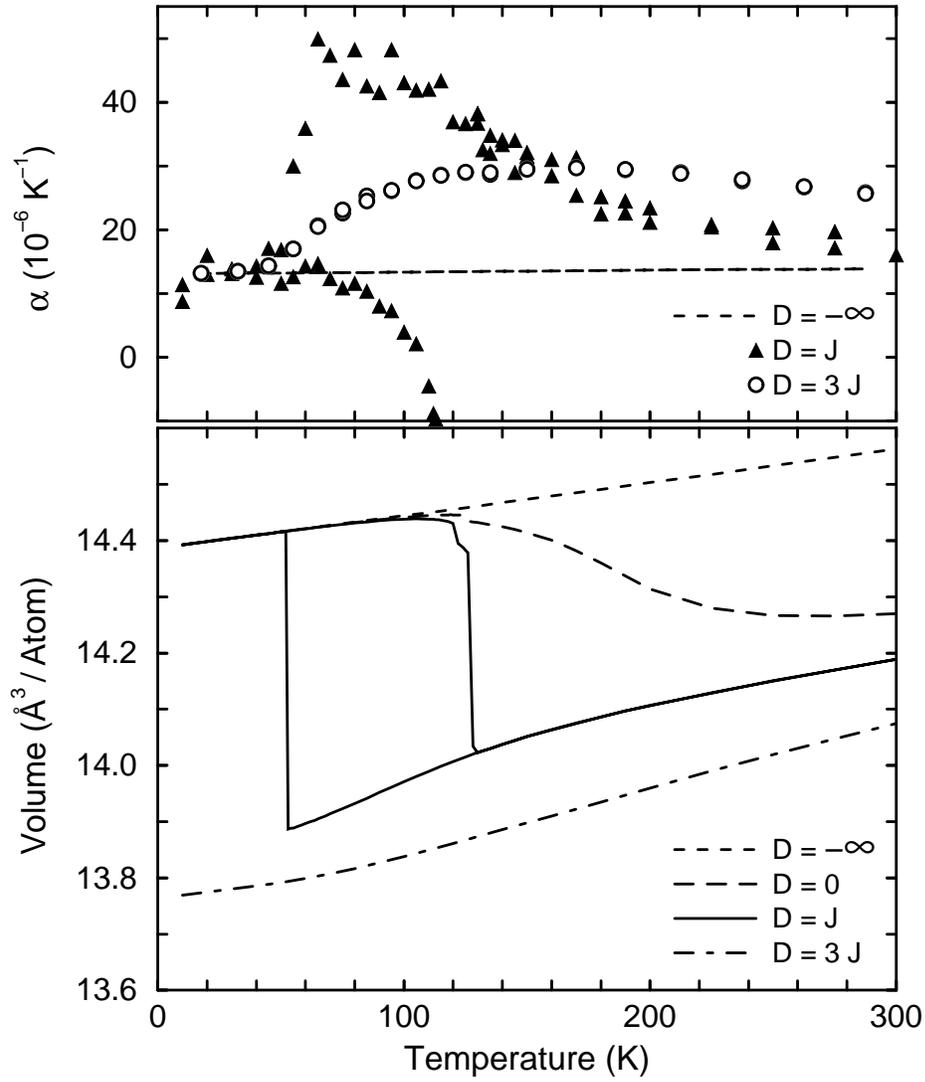}
\caption{Temperature dependence of the volume and the linear thermal
  expansion coefficient for various values of $D$.
  \label{fig:V}}
\end{figure}

Disallowing the LM spin state ($D=-\infty$), around
$T_{\rm N}=127\;{\rm K}$ a first order phase transition occurs from a type~1
AF state to a paramagnetic state.
This transition can be observed as a finite jump in the energy per
lattice site (Fig.~\ref{fig:e}).
Enlarging $D$ allows thermal excitations of the $S=0$ spin state,
leading initially to a continuous decrease of HM atoms with increasing
temperature, which appears in our results
for $D=0$ (Fig.\ \ref{fig:S2}). Due to the magnetovolume coupling 
introduced by $H_{\rm v}$,
this is connected with a continuous decrease of the volume
and, correspondingly, a reduction of the thermal expansion coefficient,
as depicted by Fig.~\ref{fig:V}.
The AF phase transition is hardly affected by the
dilution of the magnetic system
and remains of first order; only $T_{\rm N}$ is slightly reduced.
We now find an finite jump in \mean{e} at $T_{\rm N}=124\;{\rm K}$.
An investigation of the probability distribution (not shown) of the
internal energy at $T_{\rm N}$ exhibits the two maxima
structure that is typical for first order phase transitions.

For $D=J$, we find a sudden collapse of the magnetic moment at
$T_{\rm s}=127\;{\rm K}$ which is connected with an abrupt decrease of
the volume of about $3\:\%$. The magnetic phase transition takes
place a few Kelvin below at $T_{\rm N}=120\;{\rm K}$.
This situation is reflected in a double jump of the internal energy.
Upon cooling down, a large thermal hysteresis is encountered:
The system has to be cooled down to $T'_{\rm N}=53\;{\rm K}$ before it jumps
back into the AF HM state.
Responsible for this hysteresis is that lattice strain imposes a large
free energy barrier between
HM and LM states which is unlikely to be surmounted by our constant pressure
Monte Carlo algorithm.
However, the large energy barrier is suppressed in Fig.~\ref{fig:Energy}, since
at constant volume the system is forced into
a mixture of two coexisting HM and LM phases.
Similar first order transitions with a large thermal hysteresis have
also been observed in simulations of ferromagnetic systems;
for a detailed discussion we refer to \cite{FeNiPaper}.
Above $T_{\rm s}$ the mean magnetic moment is considerably lower than
the high temperature limit $\mean{S^2}=2/3$ of a spin-1 Ising model.
So with increasing temperature more atoms will be excited
to the HM state, leading to an enhanced thermal expansion.
This process is further encouraged with increasing temperature,
because the displacement of the atoms due to lattice vibrations
and the difference between HM and LM neighbor distances become comparable.
This levels out the energetic difference between 
$S_i=0$ and $S_i=\pm 1$ states introduced by $H_{\rm v}$.
For the same reason, discontinuous magnetovolume effects
do not occur for smaller values of $D$: At higher temperatures, the HM
and LM minima of the free energy are smeared out by lattice vibrations.

Above $D=2\:J$, the ground state of the system is nonmagnetic.
So HM-LM transitions as well as magnetic phase transitions
do not appear. However, we observe an enhanced thermal expansion
coefficient, due to excitation of HM atoms.

\section*{\normalsize\bf DISCUSSION}
For $D=J$, our model shows magnetovolume effects similar to
the properties of ${\rm YMn}_2$ mentioned in the introduction:
An abrupt collapse of the mean magnetic moment connected with a
decrease of the volume of several percent and a largely 
enhanced thermal expansion coefficient.
Lowering $D$ produces continuous magnetovolume effects. This resembles
the situation in ${\rm Y(}{\rm Mn}_{1-x}{\rm Al}_{x}{\rm )}_2$, where small
concentrations of Al destroy the first order transition and
anti-Invar behavior \cite{YMnAl}.
With increasing Al content the magnetovolume effects gradually vanish.
This is due to the
enlargement of the lattice constant by the Al atoms which
makes it energetically more difficult for the Mn atoms to switch to the
LM state.
On the other hand, substituting a few percent of the Y atoms
with the smaller Sc
has the opposite effect: The Mn atoms are nonmagnetic at $T=0$,
the Mn moment increases with temperature, leading to an
enhanced thermal expansion coefficient \cite{YScMn}. This is the
situation we observed for $D>2\:J$.

\subsubsection*{\normalsize\bf Acknowledgments}
This work has been supported by the
Deutsche Forschungsgemeinschaft through the SFB %Sonder\-forschungs\-bereich
166. We also thank the HLRZ (KFA J\"{u}lich) since part of the work
was done on their PARAGON parallel computer.

\end{document}